\documentclass[preprint,showpacs,preprintnumbers,amsmath,amssymb,floats,groupedaddress]{revtex4-1} 
\usepackage{graphicx}
\usepackage[english]{babel}
\usepackage{amsmath}
\usepackage{amssymb}
\usepackage{color}

\newcommand{\be}{\begin{eqnarray}} 
\newcommand{\ee}{\end{eqnarray}}

\begin{document}
\title{Thermal Hall effect in the pseudogap phase of cuprates} 
\author{Chandra M. Varma$^*$}
\affiliation{University of California, Berkeley, CA.}
\date{\today}

\begin{abstract}
The conjecture made recently by the group at Sherbrooke, that their observed anomalous thermal Hall effect 
in the pseudo-gap phase in the cuprates
is due to phonons, is supported on the basis of an earlier result that the observed loop-current order 
in this phase must  induce  lattice distortions  which are linear in the order parameter and an applied 
 magnetic field. 
The lowered symmetry of the crystal depends on the direction of the field. A consequence is that the 
elastic constants change
proportional to the field and are shown to
 induce axial thermal transport with the same symmetries as the Lorentz force enforces for the normal 
 electronic Hall effect. Direct measurements of elastic constants in a magnetic field 
 are suggested to verify the quantitative aspects of the results.
\end{abstract}
\maketitle
$^*$ Visiting Professor.

\section{Introduction}
The recent thermal Hall effect measurements in the pseudogap phase of several cuprates \cite{Taillefer-kxy2019},
\cite{Taillefer-kxy2020} 
extending to the insulating phase, join the impressive list of qualitatively new effects observed in the cuprates. 
The authors have concluded  that the heat current is most
likely carried by phonons. This is based on two observations: (1) the temperature dependence is quite unlike the 
usual relation (Wiedemann-Franz law) to the 
electrical Hall effect, and (2)
 The thermal Hall conductivity for a heat current normal to the CuO$_2$ planes
  (magnetic field parallel to the planes), $\kappa_{zy}$, 
is comparable in magnitude to the thermal Hall conductivity for a heat current 
parallel to the CuO$_2$ planes (magnetic field normal to the planes), $\kappa_{xy}$, 
as are the phonon-dominated longitudinal thermal conductivities $\kappa_{xx}$ and $\kappa_{zz}$.
 Neither electrons nor any propagating collective modes
 specific to the layered structure of the cuprates are therefore implicated. The earlier observation of $\kappa_{xy}$ 
 led to a great deal of interest  and imaginative theoretical speculations (See \cite{DHLee-JC2019} for references), which the 
 new measurements have rendered moot. The aim of this paper
 is to support the conjecture made  
\cite{Taillefer-kxy2020} 
 that the effects are due to phonons - they are really a corollary to the result \cite{Shekhter2009} derived 
 a decade ago that the order parameter predicted for the pseudo-gap phase ending at a quantum-critical point 
 \cite{cmv1997, simon-cmv} must induce a
 crystalline distortion linear in the magnetic field. 
 
 \begin{figure}[h]
\includegraphics[width=1.1\columnwidth]{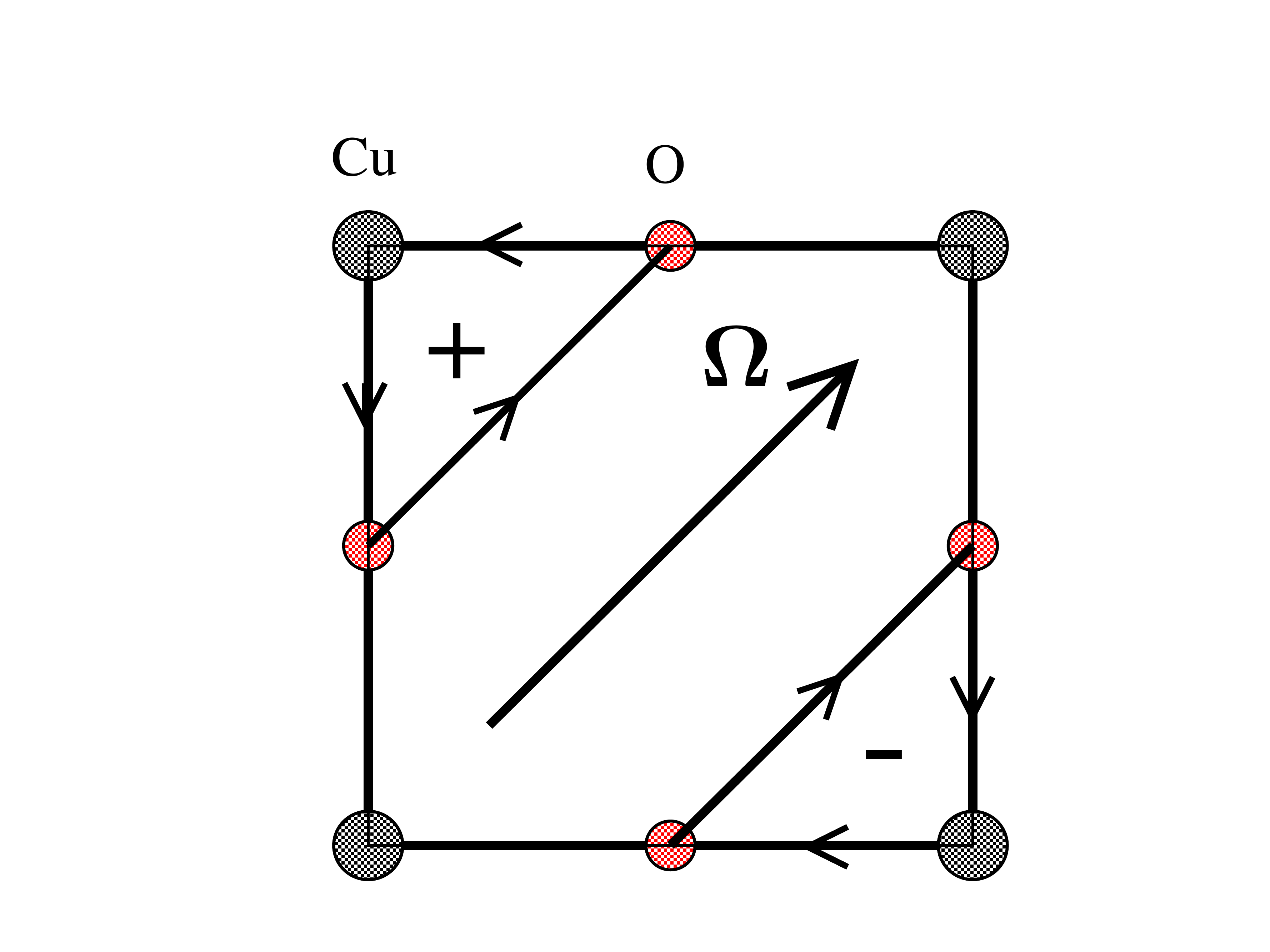}
\caption{Order parameter $\Omega$ for one its four possible orientations in the pseudo-gap phase of the
cuprates. Each unit-cell has two current loops in opposite directions so that there is no average moment in the unit-cell.
$\Omega$ is called a toroidal vector or an anapole and describes a magneto-electric order
}
 \label{Fig:OP}
\end{figure}
 
 The anomalous thermal Hall effect in cuprates is found only for temperature $T$ below the pseudogap temperature $T^*(p)$, the line 
 marking the transition to the
 pseudogap phase and increases rapidly below $T^*(p)$. The only observed symmetry change at $T^*(p)$ is the one 
 that was predicted to be to a phase of orbital currents which is odd in time-reversal, inversion and some reflections. 
 The order parameter is exhibited in Fig. (\ref{Fig:OP}).
 One or the other or several of the  aspects of altered symmetry are in evidence in a variety of different experiments \cite{Fauque2006},
 \cite{Kaminski-diARPES},
  \cite{Leridon}, \cite{Hsieh2017}, \cite{Kapitulnik1}, \cite{Armitage-Biref},  \cite{Shu2018}, \cite{Shekhter2013}, 
  \cite{Matsuda-torque1}, \cite{Matsuda-torque2}, \cite{Cerne2019} in many different cuprates, most 
 completely by polarized neutron diffraction \cite{Bourges-rev}.
 It is therefore natural to ask if the observed thermal Hall effect follows from the same symmetries. I show here that it does 
 and indeed is one of the strongest indicators of the symmetry of the pseudogap phase as it can only occur if the pseudo-gap phase is
  time-reversal and inversion odd, as specified by a toroidal and magneto-electric order parameter. 
  
  Elementary symmetry considerations 
  lead to the prediction that certain inversion odd lattice symmetry 
 changes must occur with a magnetic field applied
 in the pseudogap phase if it has the predicted order parameter.  A phonon thermal Hall requires that phonons propagating in the erstwhile symmetry directions acquire
 asymmetric bi-referingence related to the direction of the magnetic field so that the thermal current propagated by them
  flows also in the direction orthogonal to that without
 the distortion. In other words, an off-diagonal
 component, clockwise or anti-clockwise, of the energy-current tensor must develop  linearly in a field and with the
 proper antisymmetric Onsager property.
  With the changes in the lattice symmetry induced (which must
 change direction for change in direction of the magnetic field), the elastic constants acquire additional
 anisotropic elastic constants which lead to the required bi-referingence in propagation of sound and 
 of energy transport by phonons.

 \section{Lattice distortions with loop-current order \\ and magnetic field}

  I will first recollect the conclusions of Sec.IV of Ref. \cite{Shekhter2009} about distortions linear in an applied field. I will use the 
 notation ${\bf \Omega}$ for the order parameter parameter rather than ${\bf L}$ used in Ref. \cite{Shekhter2009}. 
 Consider for simplicity a tetragonal crystal, i.e. belonging to the class $D_{4h}$; lower symmetry crystals retain the distortions
  enumerated below together with smaller
 additional distortions. The magnetic field ${\bf B}$ is applied in either the $\hat{z}$-direction or the 
 ${\hat x'} \equiv (\hat{x}+\hat{y})/\sqrt{2}$ or  ${\hat y'} \equiv (\hat{x}-\hat{y})/\sqrt{2}$ directions. The order parameter
  ${\bf\Omega}$
 is either in the ${\hat x'}$ or ${\hat y'}$ directions. 
  
 A free-energy scalars is constructed from the tensor product of the distortion ${\bf u}$, the magnetic field ${\bf B}$ and
 the order parameter ${\bf \Omega}$. This is possible because ${\bf B}$ and ${\bf \Omega}$ are odd in time-inversion and
 ${\bf \Omega}$ is also odd in inversion. Then an odd in inversion distortion ${\bf u}$  is mandated. The symmetry of the distortion can be
 read from generating the irreducible representation in the $D_{4h}$ group of the product of the representations of ${\bf \Omega}$ 
 and of ${\bf B}$. Or else, by simple physical argument as follows: We note that the axial vector
 ${\bf B}_z$ transforms as $A_{2g}$ or algebraically as $i (x\partial/\partial y -  y\partial/\partial x)$ or equivalently as $ i xy(x^2-y^2)$, 
 ${\bf \Omega}_{x',y'}$
  transforms as $(i x'),( i y')$. Therefore for this orientation of the field ${\bf u}$ must transform as
 a polar vector ${\bf x'}$ or ${\bf y'}$, i.e. as $E_u$. A distortion of this form changes the rotation symmetry about the
 c-axis from four-fold to two-fold with a mirror plane going through the c-axis and the $x'$ or $y'$ axis. The point group representation
 is then $C_{2v}$. This is listed in the last column and first row of Table I. 
  
 We can similarly consider magnetic fields in the plane. Now there are many possible  induced distortions including the triclinic. With
 similar reasoning as above, the results are presented in the last four rows  in Table I, with the symmetry of the 
 distorted crystal listed in the last column.

\begin{table}[h]
\centering 
\begin{tabular}{|c|c|c|c|c|}
\hline
    {\bf Mag. field $\times$ Order parameter} &         {\bf Algebraic form of ${\mathbf u}$}       &           {\bf Irred. rep. }   &   {\bf Symmetry of distorted lattice}\\ \hline
 $\mathbf{\Omega}_{x',y'} \mathbf{B}_{z}$    &   $(x'),(y')$    &        $E_u$  &   Monoclinic($C_{2v}$) \\ \hline
 $\mathbf{\Omega}_{y'}\mathbf{B}_{y'} + \mathbf{\Omega}_{x'}\mathbf{B}_{x'}$ &  $xyz(x^2-y^2)$ &  $A_{1u}$     &                   Triclinic ($S_2$) \\ \hline
 $ \mathbf{\Omega}_{y'}\mathbf{B}_{y'} - \mathbf{\Omega}_{x'}{\mathbf{B}}_{x'} $   & $ xyz$ & $B_{1u}$ &                                     Triclinic ($C_1$) \\ \hline
$ {\mathbf{\Omega}}_{y'}{\mathbf{B}}_{x'} - {\mathbf{\Omega}}_{x'}{\mathbf{B}}_{y'}  $  & $z$   &  $A_{2u}$          &                                Tetragonal ($C_{4v}$) \\ \hline
$ {\mathbf{\Omega}}_{y'}{\mathbf{B}}_{x'} + {\mathbf{\Omega}}_{x'}{\mathbf{B}}_{y'}$  &   $z(x^2-y^2)$  & $B_{2u}$       &                       Orthorhombic ($D_2$) \\ \hline
\end{tabular}
\caption{Table of arguments and conclusions for the distortion induced by a magnetic field in the ${\hat{z}}$-direction in the loop-ordered
phase of a tetragonal crystal.}
\label{Table}
\end{table}

\section{Thermal Hall effect}
To deduce the thermal conductivity tensor due to the distortions, one must consider the elastic constants in the distorted phase.
The elastic constants change corresponding to the distortions. The elastic constants $c_{\mu, \sigma, \nu, \tau}$ are fourth-rank tensors,
defined by the equation of motion (\ref{we}) below.  There are only 6 independent elastic constants in the $D_{4h}$ symmetry
 and the resulting phonon eigen-vectors in the symmetry directions preserve the direction of propagation. In a mono-clinic
crystal, there are 13 independent elastic constants, and as we will see below they are effectively bi-refringent for the heat-current direction
relevant to the experiments. In the tri-clinic crystal, all 21 possible elastic constants are non-zero and principal axes for propagation 
cannot even 
be defined.

The elastic displacements ${\bf u}({\bf r},t)$ in a crystal
\be
{\bf u}({\bf r},t) = {\bf \epsilon}~ e^{i({\bf q}\cdot{\bf r} - \omega t)} 
\ee
 follow equations of motion 
\be
\label{we}
\rho \omega^2 {\bf \epsilon}_{\mu, \sigma} = \sum_{\tau} \sum_{\sigma, \nu} c_{\mu, \sigma, \nu, \tau} q_{\sigma}q_{\nu}~{\bf \epsilon}_{\tau, \nu}.
\ee
$\rho$ is the mass density, ${\bf q}$ the momenta, and ${\bf \epsilon}$ - the polarization vectors specify the direction and relative 
magnitude of the displacements. The eigenvalues $\omega^2$ are functions of the momenta and the polarizations. 
Thermal conductivity is given by the energy-current correlation function in the limit of long-wave-length 
and zero-frequency \cite{Luttinger1964}:
\be
\label{kappa}
\kappa_{\mu \sigma} & = & \frac{k_B \beta}{3 V} \int_0^{\infty} dt~ \int_0^{\beta} d\lambda~ <{\bf j}_{E,\mu}(0){\bf j}_{E,\sigma}(t+i\lambda)>; \\
\label{current}
~~{\bf j}_{E,\mu} & = & \sum_{{\bf q}, \tau} \frac{1}{2} \frac{\partial \omega^2_{{\bf q}, \tau}}{\partial {q_{\mu}}} n(\omega_{\bf q, \tau}/T).
\ee
${\bf j}_{E,\mu}$ is the energy-current due to the group velocity in the $\mu$-direction
of any thermally excited 
phonons; $n(\omega/T)$ is the occupation number of phonon of energy $\omega$ at temperature
$T$.

Assuming that the scattering is isotropic, 
 Eqs. (\ref{kappa}, \ref{current}, \ref{we}) state that the anisotropy of the thermal conductivity tensor  
 depends on the sum-over polarization 
of the appropriate 
 contractions of the product of two  elasticity tensors. To linear order in 
 ${\bf B}$, the off-diagonal component of $\kappa$ may be calculated by taking for one of
 the energy-currents the elastic tensors for the un-distorted tetragonal crystal, and for the other
 the elastic constants proportional to ${\bf B}$ of the distorted crystal symmetry.
 
  In general, the modes mixed in their propagation direction by the off-diagonal terms in the lowered symmetry 
  also have different polarizations, as in the examples given below. The unperturbed elastic constants 
  and their differences are 
  assumed
  to be much larger than the field induced mixing terms. So the change in 
  energies of the modes (and their attenuation) are negligible. The directions of propagation is
   determined by the perturbed eigenvectors which 
  are linearly proportional to the magnetic field, except for accidental degeneracy among the 
  unperturbed modes (of different polarizations) which will
  hardly affect the thermal conductivity tensor. The movement of current for given direction 
  of magnetic field
  is either clockwise or anti-clockwise depending on the sign of the distortion.
 All this is shown by  specific calculations below.

 \subsection{Magnetic field in ${\hat z}$-direction}
  Consider field in the ${\hat{z}}$-direction, with thermal gradient in the ${\hat{x}}$-direction. The magnetic field induced distortion
 makes the crystal monoclinic. The monoclinic crystal has 
 among its non-zero elastic constants \cite{LL-E} 
  $c_{xxxy}$ which couples a longitudinal mode traveling in the x-direction of the tetragonal crystal to 
 the transverse mode propagating in the y-direction. There is also of-course the elastic constant $c_{yxyy}$
 which couples the transverse mode propagating in the x-direction to the longitudinal mode in 
 the y-direction. These are the only possibilities of turning the heat current in the x-direction 
 to y-direction
 in a monoclinic crystal. (We need not concern ourselves with change of heat current
in the x-direction because that is only a small correction to that in the tetragonal crystal.) 
It is important that in general $c_{yxyy} \ne c_{xxxy}$. Moreover, in a
  monoclinic crystal, with appropriate choice of the
 axes, one can always take one of the elastic constants to be 0, and specify the angle between the
 positive x and y axes, say some acute value. Let us choose the x-axis to be along the a-axis of the monoclinic
 crystal and take $c_{xxxy}$ if it is the smaller of the two to be zero for magnetic field in the positive z-direction. 
 The value of $c_{yxyy}$
 of-course depends on the magnitude of the angle, which is determined by the magnitude of the 
 magnetic field. The turning of the heat current in the positive x-direction then is in the positive y-direction.
This is made more explicit by the following:

  Let $\epsilon_0(q_x,T)$ 
  and  $\epsilon_0(q_y,L)$
  be the eigenvectors, unperturbed by the 
  magnetic field, for the y-polarized transverse modes propagating in the x direction and
  longitudinal modes 
  propagating in the y-direction and $\omega_0(q_x,T) < \omega_0(q_y,L)$ be their frequencies, respectively. 
  The eigenvectors get mixed so that they  have
   additional 
   components which may be seen from the eigenvalue equation (\ref{we}) to be, 
  \be
  \label{turn}
 \delta \epsilon(q_y,T) &\approx &  \frac{c_{yxyy}q_xq_y}{\omega^2_0(q_x,L) - \omega^2_0(q_x,T)} \epsilon_0(q_x,L), \\
 \delta \epsilon(q_x,L) &\approx &  - ~ \frac{c_{xxyx}q_xq_y}{\omega^2_0(q_x,L) - \omega^2_0(q_x,T)} \epsilon_0(q_y,T).
\ee
The two contributions would turn currents in opposite directions with no net Hall effect but for the fact that 
$c_{yxyy} \ne c_{xxyx}$. As mentioned above, the smaller of these can be chosen to be 0 with appropriate 
choice of the axes. 

The ratio of $\kappa_{xy}$ to $\kappa_{xx}$ may be 
 roughly estimated from (\ref{current}) and (\ref{turn}), keeping the most essential factor, to be
 \be
 \label{xtoy}
 \frac{\kappa_{xy}(B_z)}{\kappa_{xx }} \approx \frac{2 ~c_{xxyx}(B_z)}{c_{xxxx}-c_{xyxy}}.
 \ee
 
 If the direction of the field is reversed, so is the sense of the monoclinic distortion - the acute angle above
is now an obtuse angle. By appropriate choice of axes and following the calculation above
 the thermal current  in the x-direction will
turn to the negative y-direction. Similarly for positive z-direction of the field, the current in the y-direction
will turn clockwise or anti-clockwise, keeping the same axiality as for current in the x-direction. 
All the symmetry properties of the Lorentz force in the usual 
electronic Hall 
effect are  therefore preserved. 

 The answer to the
question of the sign of the Hall effect (clockwise or anti-clockwise) 
depends on the sign of the coupling constant $\lambda$ 
in the contribution to
free-energy $\lambda ~{\bf u} ~{\bf \Omega}$. This is hard to determine since the reduction of free-energy
is determined neither by this term nor by the harmonic term in lattice distortion (see below) but by the 
anharmonic terms. There is also the vexing question of the effect of domains which is hard to answer
for order of the form shown in Fig. (\ref{Fig:OP}) if there is an equal distribution of all four domains. 
The modification of this order with a periodic pattern 
as suggested in Ref.{\cite{Varma-PLO2019} does not have the problem with domains because there is
a unique symmetry which is lower than in Fig. (\ref{Fig:OP}) and which has all the necessary attributes
necessary for the Hall effect discussed here (as well as other effects necessary for the experimental results
mentioned earlier.)

  \subsection{Magnetic field in the plane}
 For magnetic field in the plane, there are triclinic distortions and others as listed in the Table (\ref{Table}). 
 For a triclinic system, there are no
 principal axes for the elastic tensors.  Also three of the 21 elastic constants may be
 chosen to be zero by proper choice of the axes and using instead the angles between the three axes. 
 Two of 
 these are the axes in the erstwhile x to z and y to 
 z directions. By similar argument as above, there is a thermal Hall current for field in the plane and
 temperature gradient orthogonal to it in the plane due to coupling of the longitudinal modes in the plane to 
 the transverse modes in the z-direction. 
 Also in the orthorhombic
 $(C_{2v})$ class also has a non-zero $c_{yyyz} \ne c_{xxxz}$ which turns current in the
  $\hat{y}$ and $\hat{x}$ directions respectively  to the $\hat{z}$-direction. The rest is the similar to the
   case considered above. 
 So a thermal current in the plane turns partially to a current in the ${\hat{z}}$-direction. It can again be shown
 that the symmetries of this thermal Hall effect are the same as those due to Lorentz force for effect of electric
 and magnetic fields on charged particles. The magnitude of the effect for field in the
 plane is similar to that for field in the z-direction, because the normal thermal conductivity in plane and
 out of plane are similar and the requisite off-diagonal components of the elastic tensors in a field are
 expected to be similar as well.
 

 \subsection{Magnitude of the effect and related matters}
 The magnitude of the off-diagonal component of the thermal conductivity in the experiments \cite{Taillefer-kxy2019} 
\cite{Taillefer-kxy2020}
 at about 10 Tesla is 
 $O(10^{-3})$ of the diagonal component. This in turn requires, from Eq. (\ref{turn}) that the induced elastic constants at this field to be $O(10^{-3})$ 
the difference of the typical longitudinal and transverse elastic
 constants. One expects the relative change in the elastic constant to be similar to the ratio of the magnitude induced distortion to the 
 lattice constant $a$.
 Unlike arguments from symmetry, obtaining the magnitude of the effect is not really
 possible without direct measurements of the change in elastic constants and lattice distortions in a 
 magnetic field
 suggested below. 
 
 The magnitude of the distortion $|u|$ is given by the change in free-energy 
\be
\delta F(|u|/a) =  - \lambda\Big( \frac{|u|}{a}\Big) ~\Omega| ~|B| + \frac{1}{2} c ~\Big(\frac{|u|}{a}\Big)^2.
\ee
so that the relative distortion
$\frac{<|u|>}{a} = \lambda |\Omega| |B|/c$. Here $c$ is a typical elastic constant and $\lambda$ is the coupling energy. From experiments we know the 
magnitude of $\Omega$ to be about $0.1 \mu_B$ per unit-cell and from this the condensation energy due to loop-current order 
at low dopings can be
estimated to be several times larger than the maximum superconducting condensation 
energy of about $.01$ eV per unit-cell \cite{VarmaPNAS2015}.
But there is no simple way to estimate $\lambda$ and therefore $\frac{<u>}{a}$.  It is best to have direct experiments to observe 
the distortion and the change in elastic constants
   in a magnetic field. At $O(10^{-3})$ for fields of 10 Tesla, these are feasible experiments.

 The plausibility of the ideas here is reinforced by the fact that experimental results 
   consistent with another prediction
    about lattice distortions due to loop-current order.  In Ref. \cite{Shekhter2009}, it was shown that a lattice 
    distortion proportional to the square of the order parameter (for zero applied magnetic field) must occur on entering the pseudogap phase if it has the symmetries of
    loop-current order. For the single-layer compound HgBa$_2$CuO$_{4+\delta}$, this is a monoclinic distortion. Through torque -magnetometry \cite{Matsuda-torque1}, 
    anisotropy consistent with such a distortion has been observed starting at the pseudo-gap temperature. For the bi-layer compound,
    $YBa_2Cu_3O_{6+\delta}$, neutron scattering has observed \cite{Bourges-PRL2017}  that the loop-current order is mutually rotated by $\pi/2$ in the two bi-layers. Then the
    distortion expected is orthorhombic. Indeed, torque magnetometry \cite{Matsuda-torque2} observes increased anisotropy consistent with such a 
    distortion, starting again at the pseudogap temperature.
    
    In the experiments, the anti-ferromagnetic insulator La$_2$CuO$_4$ also shows a thermal Hall effect with the same characteristics and
    with magnitude continuous with that on doping to the metallic state. It would therefore be worthwhile doing direct experiments
    to look for loop-current order (which has no linear coupling to anti-ferromagnetism) in this compound. We have come across this 
    situation in the insulating antiferromagnetic compound Sr$_2$IrO$_4$, in which second-harmonic experiments \cite{Hsieh-SHG-SrIr}
    and neutron scattering experiments \cite{Bourges-SrIr} are consistent
    with loop-current order, which is observed also on doping it to a metal.
    
    Yet another test of the predicted distortions is bi-referingence and change of polarization in 
    optical propagation with a magnetic field applied. In a single crystal, it is easy enough to predict the
     symmetry
    of the dielectric tensor expected, given the Table (\ref{Table}). There are also effects of 
    the order parameter 
    in the absence of a magnetic field. Experimental results consistent with the 
    expectations \cite{VarmaEPL2014} have already been observed \cite{Armitage-Biref, Cerne2019}.

    It should be re-stated that the observed symmetry breaking, which is the basis for the
    calculation presented here, cannot be all that specifies the symmetry of the pseudo-gap phase,
    though it appears necessary. To explain the phenomena of "Fermi-arcs" and small Fermi-surface
    magneto-oscillations, a periodic modulation of the loop-current order through arrangement of topological
    defects has been proposed \cite{Varma-PLO2019}. This awaits 
    experimental verification.
    
    {\it Acknowledgements}: I wish to thank G$\ddot{a}$el Grissonnanche and Louis Taillefer for discussion on the experiments 
    and discussions with Dung-Hai Lee on aspects of the thermal Hall effects.

%

\end{document}